# SELF-CONSISTENT RESONANCE IN A PLASMA

**Evangelos Chaliasos**

365 Thebes Street

GR-12241 Aegaleo

Athens, Greece

*Abstract*

As an application of the solution of the equations of electromagnetic self-consistency in a plasma, found in a previous paper, the study of controlled thermo-nuclear fusion is undertaken. This study utilizes the resonance which can be developed in the plasma, as indicated by the above solution, and is based to an analysis of the underlying forced oscillation under friction. As a consequence, we find that, in this way, controlled thermonuclear fusion seems now to be feasible in principle. The treatment is rather elementary, and it may serve as a guide for more detailed calculations.

## 1. Introduction and the basic idea

The solution of the equations of electromagnetic self-consistency in a plasma, mentioned below in section 2, may help in heating a homogeneous plasma in a plasma reactor until the plasma reaches the high temperatures necessary for fusion nuclear reactions to take place. Up to now the heating is based mainly on the pinch effect. In this way it is possible to adiabatically compress inward the plasma, which collapses toward the axis of the cylindrical plasma column involved. The temperatures so reached are indeed high enough for fusion nuclear reactions to begin, but this mechanism contains in itself a great disadvantage: it lasts only for an extremely short interval of time. Exactly this can be overcome by the method described below.

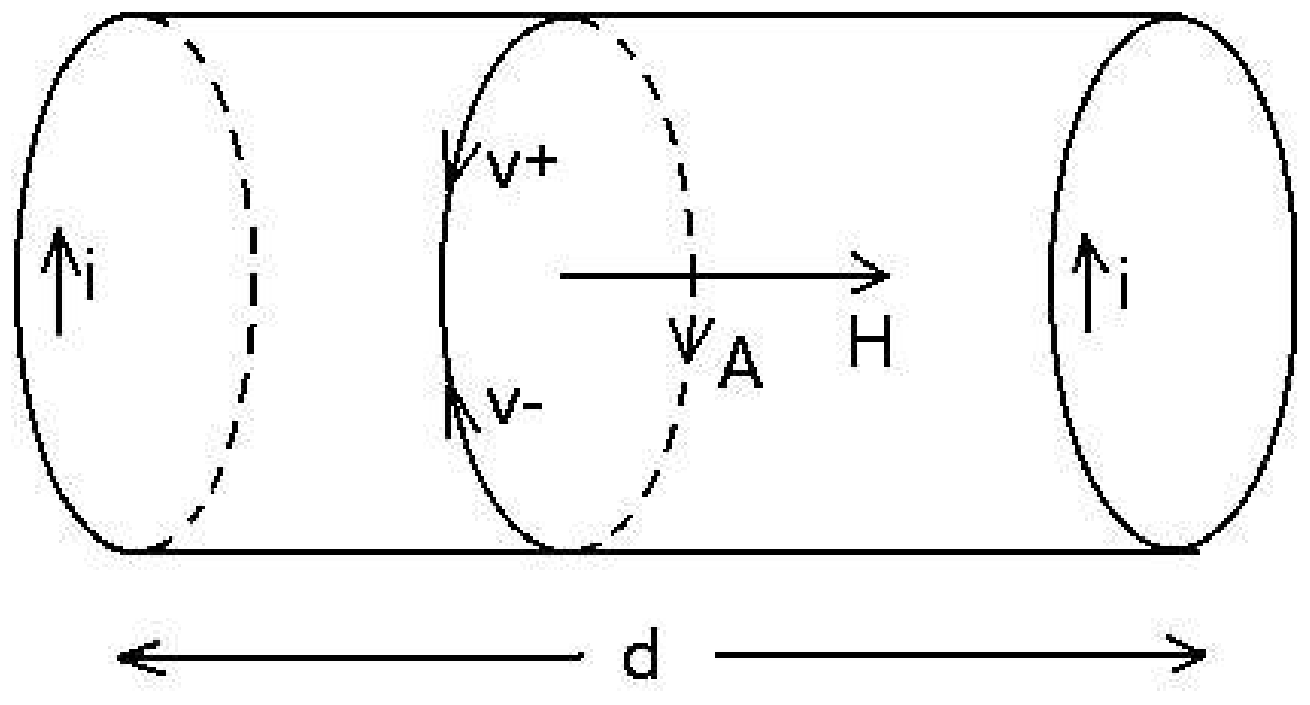

Fig.1

If we are given a cylindrical column of (homogeneous) plasma of length d, as in Fig.1, we place two spirals at its ends, and we supply them with two alternating currents i & i of the same direction, as in the figure, of suitable (cyclic) frequency ω, such that standing waves of **H**, with nodes at the ends and antinode in the middle, are created. Thus resonance results in the space of the plasma, between the spirals, concerning the field intensity **H**. This intensity results of course from a vector potential **A**, which of course will be also in resonance inside the column. Because

$$\vec{H} = curl\vec{A}, \tag{1.1}$$

and because of the positions of the spirals, **H** and **A** will be as in the figure.

Then, because of (29) and (44) of ref. [7], the plasma can be said to be "excited" and two streams of particles (namely ions and electrons) of opposite velocities will be generated, which both will move in circles rotationally oscillating in this way about the axis of the cylinder (see Fig.1), since their velocities $\mathbf{v}_+$ and $\mathbf{v}_-$ are collinear to **A**. Thus, collisions of electrons with ions (protons) will take place continuously, resulting in heating the plasma inside the column.

Equating the right hand sides of formulae (43) and (45) of ref. [7], we can determine the resonance frequency at which the resonance takes place, by setting $k \equiv 2\pi/\lambda = \pi/d$ ($d = \lambda/2$). The result is

$$\omega = c\sqrt{\frac{\pi^2}{d^2} + \frac{4\pi}{c^2}(\lambda_+ - \lambda_-)\rho} \; . \tag{1.2}$$

This frequency falls, for usual dimensions ($\lambda \approx 1$m), in the limits of radio-waves with micro-waves.

## 2. The self-consistent equations & their solution

The equations of electromagnetic self-consistency in a (two-component) plasma are [6]

$$\frac{c}{4\pi}\frac{\partial^2 A^l}{\partial x_i \partial x^i} = \sum_{+,-}\left( j_\pm{}^l \right) \quad \text{(Maxwell equations)} \tag{2.1}$$

$$\frac{1}{c}\left( A_i \frac{\partial n_\pm}{\partial x^k} - n_\pm \frac{\partial A_k}{\partial x^i} \right) u_\pm{}^i + \frac{c}{\lambda_\pm}\left( \frac{\partial n_\pm}{\partial x^k} - n_\pm u_\pm{}^i \frac{\partial u_{\pm k}}{\partial x^i} \right) = 0 \quad \text{(equation of motion)} \qquad (2.2)$$

$$\frac{\partial A^k}{\partial x^k} = 0 \quad (\mathrm{k} = 0, 1, 2, 3) \quad \text{(Lorentz condition)} \qquad (2.3)$$

$$\frac{\partial j_\pm{}^i}{\partial x^i} = 0 \quad \text{(continuity equation)} \qquad (2.4)$$

where the scalar charge density $n_\pm$ and the four-velocity $u^i_\pm$ are related to the current density $j^i_\pm$ by the relation

$$j_\pm{}^i = c n_\pm u_\pm{}^i . \qquad (2.5)$$

The solution of the above equations, for a non-relativistic homogeneous plasma, is as follows [7].

$$\vec{A} = \vec{a} \exp\{-i(\omega\ t - \vec{k} \cdot \vec{r})\} \qquad (2.6)$$

concerning the vector potential ($\varphi = 0$). Also

$$n = \rho\ = \text{ constant} \qquad (2.7)$$

and

$$\vec{v}_\pm = -(\lambda_\pm / c)\vec{a} \exp\{-i(\omega\ t - \vec{k} \cdot \vec{r})\}, \qquad (2.8)$$

or, instead of (2.8),

$$\vec{j}_\pm = \mp(\rho\ \lambda_\pm / c)\vec{a} \exp\{-i(\omega\ t - \vec{k} \cdot \vec{r})\}. \qquad (2.9)$$

In relativistic form, we have, instead of (2.8),

$$u_\pm{}^\alpha = -\frac{\lambda_\pm}{c^2} \vec{a} \exp\left\{ -i\left( \omega\ t - \vec{k} \cdot \vec{r} \right) \right\} \qquad (2.10)$$

and

$$u_{\pm}{}^{0} = -\frac{\lambda_{\pm}}{c^{2}} C_{\pm} \cong 1, \tag{2.11}$$

while, instead of (2.9),

$$j_{\pm}{}^{\alpha} = -n_{\pm} \frac{\lambda_{\pm}}{c} \vec{a} \exp\left\{-i\left(\omega\, t - \vec{k}\cdot\vec{r}\right)\right\} \tag{2.12}$$

and

$$j_{\pm}{}^{0} \equiv \pm\rho\, c = -n_{\pm} \frac{\lambda_{\pm}}{c} C_{\pm} \cong n_{\pm} c. \tag{2.13}$$

## 3. Description of the potential & the external force

We will use for convenience cylindrical coordinates (ρ, φ, z), taking as the z-axis the axis of the cylinder and the origin of coordinates in the middle between the spirals (see Fig.2). Because **H** is almost in the z-direction between the spirals, the vector potential **A** (the scalar potential will be φ = 0, since we work in the Coulomb gauge) will be of the form

$$\vec{A} = \left(0,\ A_{\varphi}\ ,0\right), \tag{3.1}$$

so that

$$\left(curl\vec{A} =\right) \frac{1}{\rho}\left[\frac{\partial}{\partial\rho}\left(\rho\ \mathrm{A}_{\varphi}\right)\hat{e}_z - \frac{\partial}{\partial z}\left(\rho\ \mathrm{A}_{\varphi}\right)\hat{e}_{\rho}\right] = \vec{H}. \tag{3.2}$$

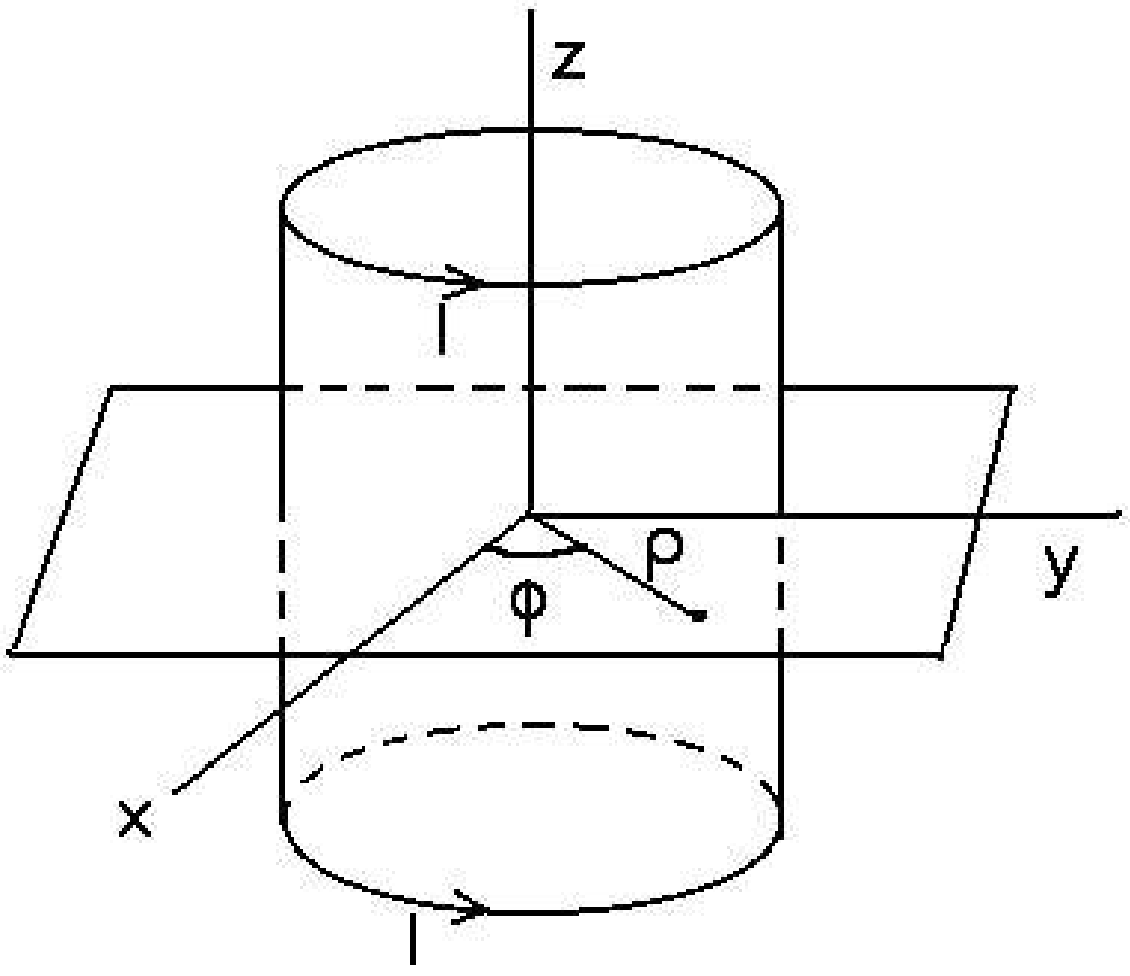

Fig.2

If we apply an alternating voltage to the spirals as to have the current

$$I = I_0 \sin\omega\ t \tag{3.3}$$

in each of them, then (we use the Gauss system throughout) the magnetic field due to one of them (the one on the left in Fig.1) will be

$$\vec{H}(z,t) = -\frac{2\pi}{Rc} I_0 \cos(\omega\ t - kz)\hat{e}_z, \tag{3.4}$$

where R is the radius of the spirals (of the cylinder).

Then, from eqns. (3.2) and (3.4), we will have for $\mathbf{A}_{\varphi}$ the equation

$$\frac{1}{\rho}\frac{\partial}{\partial\rho}\left(\rho A_\varphi\right) = -\frac{2\pi}{Rc} I_0 \cos(\omega t - kz), \tag{3.5}$$

from which we find*

$$A_\varphi(\rho, z, t) = -\frac{\pi}{Rc} I_0 \cos(\omega t - kz)\rho . \tag{3.6}$$

In order to have a resonance, ω must be taken in such a way that standing waves of $\mathbf{A}_\varphi$ are formed between the two spirals. Thus, if we incorporate the current of the other spiral as well, we must have

$$A_\varphi(\rho, z, t) = -\frac{\pi}{Rc} I_0 \cos(\omega t - kz)\rho - \frac{\pi}{Rc} I_0 \cos(\omega t + kz)\rho , \tag{3.7}$$

resulting in

$$A_\varphi(\rho, z, t) = -\frac{2\pi}{Rc} I_0 \rho \cos kz \cos\omega t. \tag{3.8}$$

Then, the velocity of the organized motion of protons and electrons will be (cf. eqns. (2.6) and (2.8))

$$\vec{V}_\pm = -\frac{\lambda_\pm}{c}\vec{A}. \tag{3.9}$$

The force acting on these particles is

$$\vec{f}_\pm \equiv m_\pm \dot{\vec{V}}_\pm = -\frac{\pm e}{c}\left(0, \dot{A}_\varphi, 0\right) \left(= \pm e\vec{E}\right). \tag{3.10}$$

Thus, since

$$\dot{A}_\varphi = \omega \frac{2\pi}{Rc} I_0 \rho \cos kz \sin\omega t, \tag{3.11}$$

we get

$$f_{\varphi\pm} = -(\pm e)\frac{2\pi}{Rc^2}\omega\rho I_0 \cos kz \sin\omega t. \tag{3.12}$$

It is obvious that

$$f_{\varphi +} = -f_{\varphi -}. \tag{3.13}$$

Thus, I claim that, if initially for the momenta

$$p_{+0} = -p_{-0} (\approx 0), \tag{3.14}$$

* Concerning the $e_\rho$ direction, we find simply that z/t = ω/k (phase velocity), for the motion of the wave in the z-direction.

then at all moments it will be

$$p_+ = -p_- . \tag{3.15}$$

This is indeed the case if we use the notion of the push of force. Thus

$$\Delta p_+ = \int f_{\varphi\,+} dt = -\int f_{\varphi\,-} dt = -\Delta p_- , \tag{3.16}$$

resulting in (3.15), because of (3.14). We set

$$-p_- = p_+ \equiv p. \tag{3.17}$$

If we then consider the kinetic energies of protons and electrons as being due to the corresponding "temperatures" $T'_\pm$ , we will have

$$\frac{p^2}{2m_+} = \frac{1}{2} kT'_+ \quad \& \quad \frac{p^2}{2m_-} = \frac{1}{2} kT'_- , \tag{3.18}$$

from which

$$\frac{T'_-}{T'_+} = \frac{m_+}{m_-} , \tag{3.19}$$

or

$$m_+ T'_+ = m_- T'_- . \tag{3.20}$$

We will make use of these relations in what follows.

## 4. The internal friction coefficient

The internal friction in the plasma is due to the relative (organized) motion of protons and electrons under the influence of the oscillating potential. We have for the velocity of protons with respect to electrons

$$V_{rel} = V_+ - V_- . \tag{4.1}$$

Because of eqn (3.9), we obtain then

$$V_{rel} = \left(1 + \frac{m_+}{m_-}\right) V_+ \left(\cong 1837 V_+\right) \tag{4.2}$$

Similarly, for the velocity of electrons with respect to protons,

$$-V_{rel} = V_- - V_+, \tag{4.3}$$

we get

$$-V_{rel} = \left(1 + \frac{m_-}{m_+}\right) V_-, \tag{4.4}$$

so that we can write for both eqns. (4.2) and (4.4) simply

$$\pm V_{rel} = \left(1 + \frac{m_\pm}{m_\mp}\right) V_\pm \tag{4.5}$$

Now, concerning the coefficient of internal friction (the "damping coefficient") $p_\pm$ we have (see Fig.3) for the force

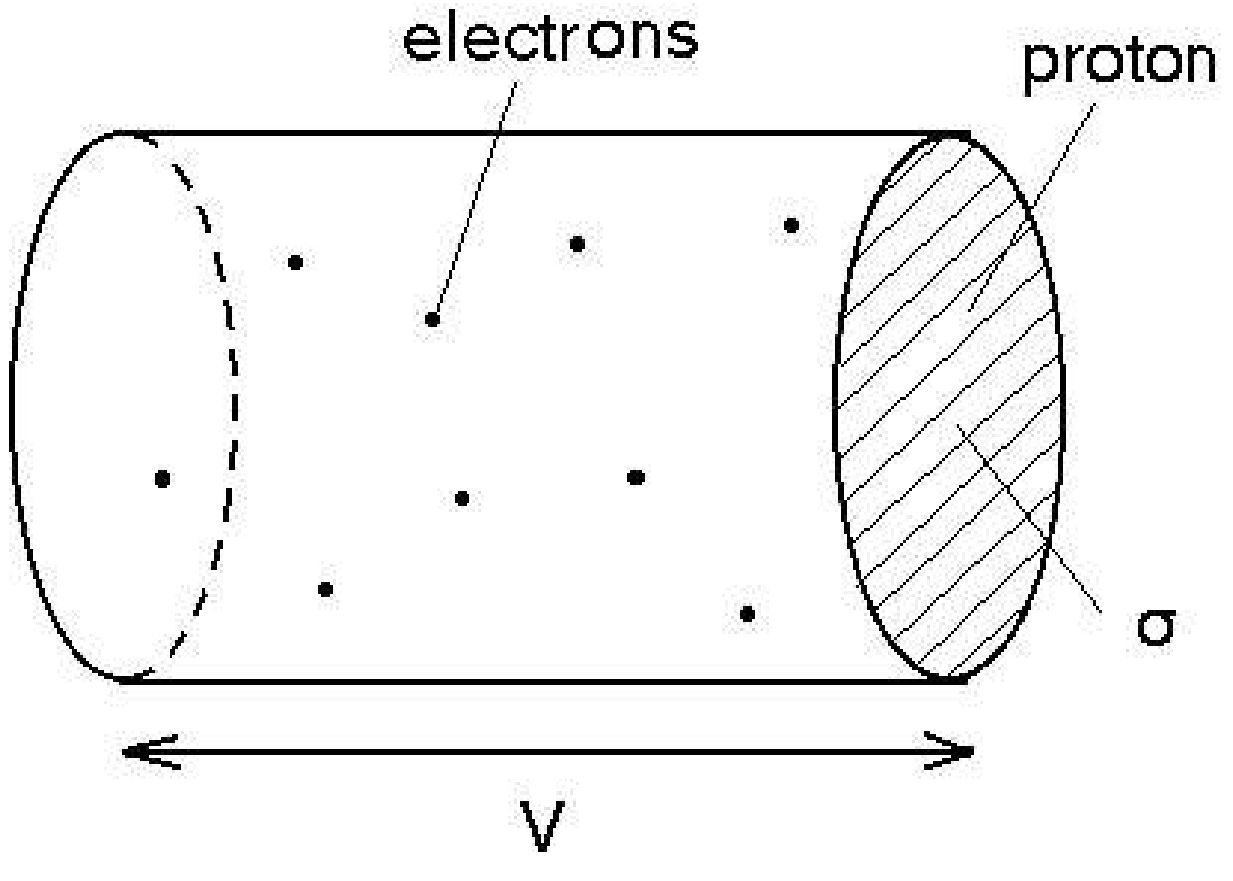

Fig.3

$$F_{friction} = (\sigma_{pe} V_{rel,rms} N)(m_\pm V_{rel}), \tag{4.6}$$

where σpe is the cross-section for p-e collisions and N the number density of protons or electrons, so that

$$\pm p_{\pm} = \sigma_{pe} V_{rel,rms} N m_{\pm}. \qquad (4.6')$$

Thus we find

$$p_{\pm} = \pm\sigma_{pe}\left(1+\frac{m_{\pm}}{m_{\mp}}\right)V_{\pm rms} m_{\pm} N. \qquad (4.7)$$

But

$$\frac{1}{2}m_{\pm}V_{\pm rms}{}^{2}\ \left(=\frac{1}{2}m_{\pm} < V_{\pm}^{2} >\right) = \frac{1}{2}kT'_{\pm}\ , \qquad (4.8)$$

from which we get

$$V_{\pm rms} = \pm\sqrt{\frac{kT'_{\pm}}{m_{\pm}}}, \qquad (4.9)$$

so that finally

$$p_{\pm} = \sigma_{pe}\left(1+\frac{m_{\pm}}{m_{\mp}}\right)\sqrt{m_{\pm}kT'_{\pm}}\,N \qquad (4.10)$$

## 5. Equation of motion and its solution

The equation of motion for protons (+) and electrons (-) under the influence of the external force $f_{\varphi\pm}$ and under friction with damping coefficient $p_{\pm}$ is

$$m_{\pm}\ddot{\varphi}_{\pm} + p_{\pm}\dot{\varphi}_{\pm} + m_{\pm}\omega_0^{2}\varphi_{\pm} = f_{\varphi\pm}\ , \qquad (5.1)$$

where $\omega_0$ is the natural frequency given by (1.2), or, substituting for $f_{\varphi\pm}$ and $p_{\pm}$ ,

$$m_{\pm}\ddot{\varphi}_{\pm} + \sigma_{pe}\left(1+\frac{m_{\pm}}{m_{\mp}}\right)\sqrt{m_{\pm}kT'_{\pm}}\,N\dot{\varphi}_{\pm} + m_{\pm}\omega_0^{2}\varphi_{\pm} =$$

$$= -(\pm e)\frac{2\pi}{Rc^{2}}\omega\, I_0 \cos kz \sin\omega\, t \qquad (5.2)$$

For the steady part of the solution (after the settlement effects), we have [1]

$$\varphi_{\pm} = \frac{\mp \frac{2\pi e}{Rc^2} \omega I_0 \cos kz}{\sqrt{m_{\pm}^2 \left(\omega_0^2 - \omega^2\right)^2 + p_{\pm}^2 \omega^2}} \sin\left(\omega t - \varphi\right), \tag{5.3}$$

or,

$$\varphi_{\pm} = \varphi_{\pm 0} \sin(\omega t - \varphi) \tag{5.4}$$

with

$$\varphi_{\pm 0} = \frac{\mp \frac{2\pi e}{Rc^2} \omega I_0 \cos kz}{\sqrt{m_{\pm}^2 \left(\omega_0^2 - \omega^2\right)^2 + p_{\pm}^2 \omega^2}} \tag{5.5}$$

and

$$\tan\varphi = \frac{p_{\pm}\omega}{m_{\pm}\left(\omega_0^2 - \omega^2\right)}. \tag{5.6}$$

At resonance $\omega = \omega_0$. Then $\tan\varphi = \infty$, so that $\varphi = \pi/2$ and

$$\varphi_{\pm 0} = \mp \frac{2\pi e I_0}{Rc^2 p_{\pm}} \cos kz. \tag{5.7}$$

Thus

$$\varphi_{\pm} = -\varphi_{\pm 0} \cos\omega_0 t, \tag{5.8}$$

and

$$\varphi_{\pm rms} = \frac{\left|\varphi_{\pm 0}\right|}{\sqrt{2}}. \tag{5.9}$$

We also have, upon differentiation,

$$\dot{\varphi}_{\pm} = \omega_0 \varphi_{\pm 0} \sin\omega_0 t, \tag{5.10}$$

and

$$\dot{\varphi}_{\pm rms} = \omega_0 \frac{\left|\varphi_{\pm 0}\right|}{\sqrt{2}}, \tag{5.11}$$

so that, from

$$V_{\pm} = \rho \dot{\varphi}_{\pm} \tag{5.12}$$

we get

$$V_{\pm rms}\left(= \rho \dot{\varphi}_{\pm rms}\right) = \rho \omega_0 \left|\varphi_{\pm 0}\right| / \sqrt{2}. \tag{5.13}$$

But

$$\frac{1}{2}m_\pm V_{\pm rms}{}^2 = \frac{1}{2}kT'_\pm, \tag{5.14}$$

so that

$$T'_\pm = \frac{1}{k}m_\pm V_{\pm rms}{}^2, \tag{5.15}$$

from which we obtain

$$T'_\pm = \frac{1}{k}m_\pm \rho^2 \omega_0{}^2 \varphi_{\pm 0}{}^2/2, \tag{5.16}$$

and, substituting for $\varphi_{\pm 0}$ from (5.7),

$$T'_\pm = \frac{m_\pm}{2k}\rho^2\omega_0{}^2\left(\pm\frac{2\pi\, eI_0}{Rc^2 p_\pm}\cos kz\right)^2. \tag{5.17}$$

Eliminating ρ with R since they are of the same order of magnitude, and substituting for $p_\pm$ from eqn. (4.10), we get

$$T'_\pm \cong \frac{1}{k}\omega_0{}^2\frac{2\pi^2 e^2}{c^4}I_0{}^2\cos^2 kz\cdot\frac{1}{\sigma_{pe}{}^2\left(1+\frac{m_\pm}{m_\mp}\right)^2 kT'_\pm N^2}, \tag{5.18}$$

so that

$$T'^2_\pm \cong \frac{2\pi^2 e^2}{k^2c^4}\omega_0{}^2 I_0{}^2\cos^2 kz\cdot\frac{1}{N^2\sigma_{pe}{}^2\left(1+\frac{m_\pm}{m_\mp}\right)^2}, \tag{5.19}$$

from which we find

$$T'_\pm \cong \sqrt{2}\frac{\pi\, e}{kc^2}\frac{1}{\left(1+\frac{m_\pm}{m_\mp}\right)}\frac{1}{\sigma_{pe}N}\omega_0 I_0\cos kz, \tag{5.20}$$

or finally

$$T'_\pm \cong \frac{\sqrt{2}\pi}{1837}\frac{e}{kc^2}\frac{\omega_0 I_0}{\sigma_{pe}N}\cos kz. \tag{5.21}$$

Working an arithmetic example, in the Gauss system of course, we take

$e = 4.8\times10^{-10}$ ESU

$k = 1.38\times10^{-16}$ erg deg$^{-1}$

$c = 3\times10^{10}$ cm sec$^{-1}$

[4] and

$\omega_0 = 2\pi\times10^{9}$ sec$^{-1}$

$\sigma_{pe} = \pi$ (2 Fermi)$^2 = 4\pi\times10^{-26}$ cm$^2$

$N = 10^{12}$ cm$^{-3}$.

Then we find that, in order to have a « temperature » of $T'_+ = 10^7$ deg (necessary for fusion nuclear reactions to begin), we have to supply a current of $I_0 \cong 7.1$ nA. (This is comparable to currents in electronic circuits [5], [15]).

## 6. Energy consumption

If we write for the consumption $I_\pm$ at the frequency $\omega = \omega_0 + \varepsilon$, with small $\varepsilon$ (almost resonance), $I_\pm = I_\pm(\varepsilon)$, then specifically for $\varepsilon = 0$ (exact resonance) we have [13]

$$I_\pm(0) = \frac{f_\pm^{\,2}}{4m_\pm\lambda_\pm}, \tag{6.1}$$

with the meaning of the symbols as given in the reference above. Thus, in our notation,

$$f_{\pm} = \mp \rho \frac{2\pi e}{Rc^2} \omega_0 I_0 \cos kz, \tag{6.2}$$

and

$$2m_{\pm}\lambda_{\pm} = p_{\pm}, \tag{6.3}$$

$$f_{\pm}^2 = \rho^2 \frac{4\pi^2 e^2}{R^2 c^4} \omega_0^2 I_0^2 \tag{6.4}$$

so that

(for the middle between the spirals) and

$$\lambda_{\pm} = \frac{p_{\pm}}{2m_{\pm}}. \tag{6.5}$$

Introducing these $f_{\pm}^2$ and $\lambda_{\pm}$ in (6.1), eliminating again $\rho^2$ with $R^2$, and simplifying, we get

$$I_{\pm}(0) \cong \frac{\pi^2 e^2}{c^4} \omega_0^2 I_0^2 \frac{2}{\sigma_{pe} N \left(1 + \frac{m_{\pm}}{m_{\mp}}\right)\sqrt{m_{\pm} k T'_{\pm}}}, \tag{6.6}$$

or finally

$$I_{+}(0) \cong \frac{2\pi^2 e^2}{1837 c^4} \frac{\omega_0^2 I_0^2}{\sigma_{pe} N} \frac{1}{\sqrt{m_{+} k T'_{+}}} \quad \& \quad I_{-}(0) = 1836 I_{+}(0). \tag{6.7}$$

Inserting to this formula the values of our arithmetic example, and taking $m_- = 9.1\times10^{-28}$ gr [4], we find $I_+(0) \approx 9.5\times10^{-12}$ erg/sec & $I_-(0) \approx 1.7\times10^{-8}$ erg/sec. These values apply of course to only one particle, proton (+) or electron (-). For N particles ($cm^{-3}$) we find $NI_+(0) + NI_-(0) \approx 1.7\times10^5$ (erg/sec)/$cm^3$. For volume $V=1m^3$, if $I \equiv V [NI_+(0) + NI_-(0)]$, we find finally $I \approx 15$ kW/$m^3$.

## 7. Settlement time, mean free path & mean free time

The settlement time is a measure of the time required for the steady solution (particular solution) of eqn. (5.1) to be established. This is the time required for the non-steady solution (the general solution of the homogeneous equation resulting from (5.1)) to fall to $e^{-1}$ times its initial value. It equals therefore to the inverse of the damping coefficient λ, that is 1/λ [13]. Since λ may change from protons to electrons, we will have from (6.5), if we take also into account eqn. (4.10), the equation

$$\frac{1}{\lambda_{\pm}} = 2m_{\pm}\frac{1}{\sigma_{pe}N}\frac{1}{\left(1+\dfrac{m_{\pm}}{m_{\mp}}\right)\sqrt{m_{\pm}kT'_{\pm}}}. \tag{7.1}$$

From this, it can be seen that

$$\frac{1}{\lambda_{-}} = \frac{1}{\lambda_{+}} \equiv \frac{1}{\lambda} = 2\cdot\frac{1836}{1837}m_{-}\frac{1}{\sigma_{pe}N}\frac{1}{\sqrt{m_{\pm}kT'_{\pm}}}, \tag{7.2}$$

because of eqn (3.20).

With the values of our arithmetic example, we find that 1/λ ≅ 300 sec = 5 min.

The mean free path of the colliding protons and electrons, before homogenization of the velocities ("thermalization"), is [14]

$$\text{M.F.P.} = \frac{1}{\sigma_{pe}N}, \tag{7.3}$$

and the mean free time (under the same circumstances) [14]

$$\text{M.F.T.} = \frac{\text{M.F.P.}}{V_{rel,rms}} = \frac{\text{M.F.P.}}{1837V_{+rms}} = \frac{\text{M.F.P.}}{1837}\sqrt{\frac{m_{+}}{kT'_{+}}}. \tag{7.4}$$

In our arithmetic example, we find M.F.P. ≅ $8\times10^{10}$ m and M.F.T. ≅ 152 sec = 2.5 min.

We have to consider this mean free time as a measure of the thermalization time ("time of relaxation"), after which $T'_{\pm} = T_{\pm}$ (the actual temperatures of protons and electrons), since it indicates that after this time almost all protons have collided with (almost all) electron, and thus homogenization of their velocities has most likely occurred.

**8. Thermonuclear energy released**

We are now interested in collisions of protons with protons. Of course this will result in fusion, after thermalization at $T_+ \sim 10^7$ deg, a temperature required in order for fusion nuclear reactions to begin. If the cross-section for p-p fusion is $\sigma_{pp}$ and v denotes *thermal* velocity, then the number of collisions of a proton with other protons will be $\sigma_{pp} v_{+rms} N$. This (assuming $\sigma_{pp} \sim \sigma_{pe}$) results in $6.3\times10^{-6}$ $sec^{-1}$ collisions in our arithmetic example.

The rate then of p-p collisions (for a more detailed analysis see [9]) resulting in fusion will be $\sigma_{pp} v_{+rms} N^2/2$.

To find the required velocity, we must observe that

$$\frac{1}{2} m_+ {v_{+rms}}^2 = \frac{3}{2} k T_+ . \qquad (8.1)$$

Thus

$$v_{+rms} = \sqrt{\frac{3kT_+}{m_+}}, \qquad (8.2)$$

so that, in our arithmetic example, we find $v_{+rms} \cong 5\times10^7$ cm/sec.

The rate of p-p fusion collisions will be then

$$\text{Rate} = \frac{1}{2} N^2 \sigma_{pp} \sqrt{\frac{3kT_+}{m_+}}, \qquad (8.2')$$

resulting, in our arithmetic example, in $3\times10^6$ collisions $sec^{-1}$ $cm^{-3}$.

We will take as an example the simplest of the reactions involved, namely [10]

$$H^1 + H^1 \rightarrow H^2 + e^+ + \nu \quad + 0.42 \text{ MeV} \tag{8.3}$$

If we add to 0.42 MeV the energy released by the annihilation of the positron, we gain totally 1.44 MeV from each pair p-p. Thus, we find for the energy release

$$\text{Energy release} \equiv \text{Rate} \times 1.44 \text{ MeV}, \tag{8.4}$$

about 0.7 W/m$^3$.

## 9. Rate of temperature increase

After settlement has been established, the energy consumed will equal the heat produced by dissipation [13]. This will be evidently equal to the rate of increase of the energy content of the plasma. Assuming that the plasma temperature $T_\pm$ corresponds to $T'_\pm$ *after* homogenization of the velocities, we may thus write

$$I_\pm(0) = \frac{d}{dt}\left(\frac{3}{2} kT_\pm\right), \tag{9.1}$$

From which, because of eqn. (6.7),

$$\frac{2\pi^2 e^2}{1837 c^4} \frac{\omega_0^2 I_0^2}{\sigma_{pe} N} \frac{1}{\sqrt{m_+ k T'_+}} = \frac{3}{2} k \frac{dT_+}{dt}, \tag{9.2}$$

so that, after integration,

$$T_+ = \frac{4\pi^2 e^2}{3 \cdot 1837 k c^4} \frac{\omega_0^2 I_0^2}{\sigma_{pe} N} \frac{1}{\sqrt{m_+ k T'_+}} t. \tag{9.3}$$

For the arithmetic example, we have $I_+(0) \sim 9.5\times10^{-12}$ erg/sec. If we want to find the time it takes the plasma to thermalize at $T_+ = 10^7$ deg, we have to use (9.1), so that this time is

$$t = \frac{3}{2} kT_+ / I_+(0). \tag{9.4}$$

We get thus for the thermalization time t ~ 218 sec =3.6 min.

We have to observe that all times we have found (settlement time, mean free time and thermalization time) are of the same order of magnitude, that is of a few minutes.

**10. Energy content and Pressure**

The energy content of the plasma, after thermalization, due to protons will be

$$E_+ = \frac{1}{2} m_+ v_{+rms}{}^2 N = N \cdot \frac{3}{2} kT_+. \tag{10.1}$$

The same quantity due to electrons will be

$$E_- = N \cdot \frac{3}{2} kT_- = N \cdot \frac{3}{2} k \cdot 1836 T_+. \tag{10.2}$$

In our arithmetic example, for $T_+ = 10^7$ °K, we find respectively $E_+ \cong 2\times10^2$ Joule/m³ & $E_- \cong 3.8\times10^5$ Joule/m³.

Thus, for the total energy content,

$$E = E_+ + E_-, \tag{10.3}$$

we find under the same circumstances $E \cong E_- = 3.8\times10^5$ Joule/m³.

The kinetic pressure will be

$$\text{Pressure} = NkT_+ + NkT_- = 1837 NkT_+, \tag{10.4}$$

resulting, in our arithmetic example, in $2.5\times10^6$ erg/cm³.

But, if the confining *static* magnetic field is H, we will also have for the magnetic pressure

$$\text{Pressure} = H^2/8\pi. \tag{10.5}$$

Equating the two, we find in our arithmetic example that the required magnetic field is H ≈ $7.9\times10^3$ Gauss.

We use the formula

$$H = \frac{4\pi}{c} in^*, \tag{10.6}$$

where i is the direct current flowing inside the confining coil and n* is the number of spirals per unit length of the coil.

We thus find

$$in^* = \frac{c}{4\pi} H, \tag{10.7}$$

so that, in our arithmetic example, we get in* ≅ 632 kA/m.

11. **Stability** (see [13])

The equation of motion is of the form

$$m\ddot{\varphi} + p\dot{\varphi} + m\omega_0^2\varphi = f_\varphi , \tag{11.1}$$

with $f_\varphi = f_{\varphi 0}\cos\omega t$. The corresponding homogeneous equation is

$$m\ddot{\varphi} + p\dot{\varphi} + m\omega_0^2\varphi = 0. \tag{11.2}$$

The general solution of (11.2) is of the form

$$\varphi_{\text{hom}} = ae^{-\lambda t}\cos(\omega_0 t + \alpha), \tag{11.3}$$

with $\lambda = p/(2m)$, while a particular solution of (11.1) is of the form

$$\varphi_{par} = b\cos(\omega t + \delta). \tag{11.4}$$

Thus the general solution of (11.1) is

$$\varphi = \varphi_{\text{hom}} + \varphi_{par} = ae^{-\lambda t}\cos(\omega_0 t + \alpha) + b\cos(\omega t + \delta). \tag{11.5}$$

Suppose now that we set

$$\varphi'(t) = \varphi(t) + \varepsilon(t), \tag{11.6}$$

where ε(t) is a small perturbation, and we demand that (11.6) is a solution of (11.1). Then we will examine the behavior of ε(t) as t increases, namely whether it is growing or decaying. Inserting (11.6) in (11.1) instead of φ, we find that the terms in φ cancel each other by means of the unperturbed eqn. (11.1), so that we are left with ε(t) having to satisfy the homogeneous eqn. (11.2) if we place it in the position of φ. That is ε(t) has to satisfy the equation

$$m\ddot{\varepsilon} + p\dot{\varepsilon} + m\omega_0^2\varepsilon = 0. \qquad (11.7)$$

Thus the solution for ε has to be of the form, similar to (11.3),

$$\varepsilon = ae^{-\lambda t}\cos(\omega_0 t + \alpha). \qquad (11.8)$$

Thus

$$\varepsilon(t) \xrightarrow[t\to\infty]{} 0, \qquad (11.9)$$

that is ε is decaying and therefore the solution (11.5) for φ is stable.

## 12. Discussion

Of course in a fusion machine we will not have only one “cell” (with two spirals at its ends) of length d (see Fig.1), but many such cells placed successively one by the other. If the whole configuration is linear, the plasma can of course be successfully confined in the direction of ρ by the static homogeneous magnetic field due to a suitable coil surrounding the curved surface of the resulting long cylinder, carrying the appropriate direct current. There is a problem only at the two bases of the configuration. Of course we can appropriately modify the coil at the two ends of the whole cylinder so that two magnetic mirrors are formed at the two bases of the so-formed magnetic bottle. But we can by no way achieve a perfect reflection of the plasma on these two magnetic mirrors, since we will always have an even small escape of the plasma particles there. We can however "bend" this configuration and identify the two bases, so as for the cylinder to form a torus. In this way it is clear that there will be no escape at all, since now *all* of the plasma particles will move indefinitely inside the torus. It is also evident that the preceding analysis will be approximately valid concerning the torus, the better the smaller curvature along the torus, that is the better the longer the big radius of the torus is.

In the steady state of operation of the machine the energy consumed at resonance I(0) will dissipate inside the torus maintaining the plasma heated and so supporting the fusion. But it is *not* lost. We will finally take it again at the "output" of the machine[§] together with the heat resulted by the fusion. If then we use the former in order to (almost) take again the initial current I required, it is obvious that the energy released (e.r.) by the fusion will be (almost) the *net* gain of energy.

Of course the yield of the machine will be the better the larger the ratio (e.r.)/I(0) is. In our arithmetic example the yield is not so good. But we have to observe that (from eqns. (6.7) and (5.21)) $I(0) \propto N$, while (see §8) e.r. $\propto N^2$. Thus we can improve the yield by just increasing N. Then we will also have (from eqn. (5.21)) $I_0 \propto N$, while (from eqns. (10.6), (10.5) and (10.4)) $in^* \propto \sqrt{N}$. Thus increasing N does not have a strong effect on increasing $in^*$ as well. For example if we increase N by a factor of 100, we will be required to increase $in^*$ by a factor of only 10, but we will take a (thermo-nuclear) energy release increased by a factor of 10000 (please apply this to our arithmetic example).

Another issue with which we will be confronted in the design and real construction of such a machine has to do with how to stabilize the temperature of the plasma inside the reactor, at $T_+ = 10^7$ °K say, preventing it from increasing more[%]. Suppose that the plasma torus, of short radius $r_0$ and long radius R, is surrounded by a concave toroidal layer of material, of thickness l, which is a bad conductor of heat, of a coefficient of thermal conduction κ (e.g. κ = 0.0004). Suppose also that the whole system of the plasma torus with its material cover is embedded in a tank filled with water.

[§] Of course part of I(0) is converted to bremsstrahlung energy, but it is also *not* lost because it will be absorbed by the surrounding walls of the reactor, being finally converted to heat.

Let the outer surface of the plasma, of area $S_1$ , which is approximately the same with the inner surface of the surrounding material, have a temperature $T_1$. Let also the outer surface of the material, of area $S_2$ , have a temperature $T_2$, which is the same with the temperature of the water. Then the heat excess (per unit time) Q/t =[I (0)+(e.r.)]V of the plasma torus will be transferred by conduction from $S_1$ to $S_2$, being absorbed finally by the water. Then, for the heat flow (which will be the same with [I(0)+(e.r.)]V in the steady state we examine) Q/t, we will have, from the law of heat conduction [3],

$$\frac{Q}{t} = -\kappa\, S \frac{dT}{dx}, \tag{12.1}$$

where S is the area of a surface (inside the material) perpendicular to the direction of (heat) flow and dT/dx is the temperature gradient inside the material. Writing eqn. (12.1) as

$$\frac{Q}{t}\frac{dx}{S(x)} = -\kappa\, dT \tag{12.2}$$

and integrating, we find

[%] Of course $T_\pm$ cannot become greater than $T'_\pm$ , its cause, the fusion reactions kept apart (the heat excess, due to $I_\pm(0)$, must be necessarily dissipated outside the reactor). But right because of these fusion reactions, when they begin, the temperatures $T_\pm$ will undoubtedly increase, if we do not take care of it.

$$T_1 - T_2 = \frac{1}{\kappa}\frac{Q}{t}\int_0^l \frac{dx}{S(x)}, \qquad (12.3)$$

where l is the thickness of the material. Since we know $T_1$, and letting $T_2$ be the boiling temperature of the water, we can determine l, so that Q/t =[I(0)+(e.r.)]V, as follows.

We first find for

$$I(l) \equiv \int_0^l \frac{dx}{S(x)} \qquad (12.4)$$

that

$$I(l) = \frac{1}{4\pi^2 R}\ln\left(1 + \frac{l}{r_0}\right). \qquad (12.5)$$

After substituting this value in (12.3), and letting Q/t =[I(0)+(e.r.)]V, where V is the volume of the plasma torus, which we find that is given by

$$V = 2\pi^2 R r_0^2, \qquad (12.6)$$

we finally find

$$l = r_0\left\{-1 + \exp\left[\frac{2\kappa\,(T_1 - T_2)}{r_0^2\,(I(0) + (e.r.))}\right]\right\}. \qquad (12.7)$$

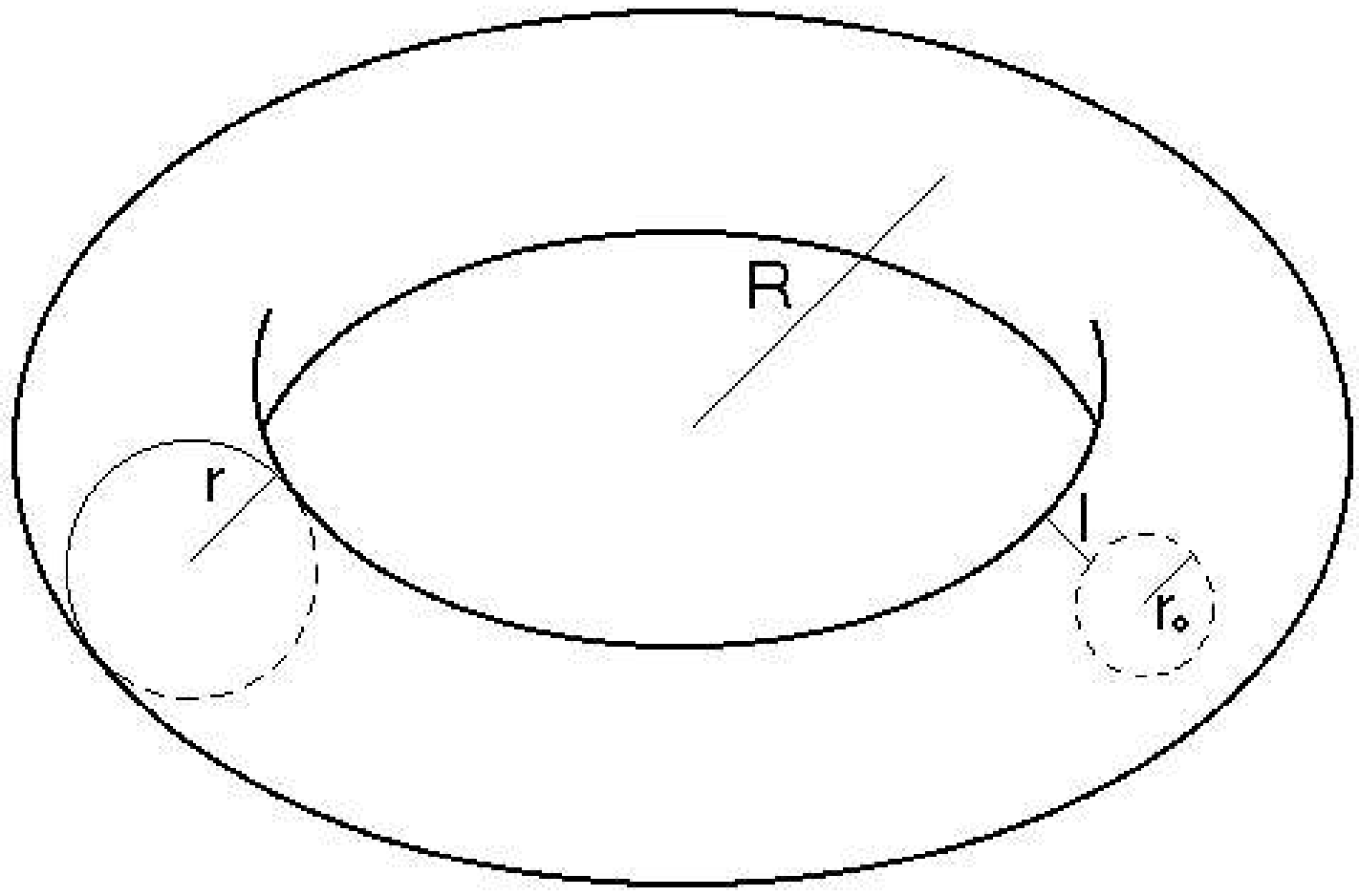

Fig.4

In our arithmetic example, take $r_0 = 1$ m, $T_1$-$T_2 \sim 10^7$ °K, and $\kappa = 0.0004$ cal sec$^{-1}$ cm$^{-1}$ grad$^{-1}$ (1cal ≅ 4 Joule). We find that, if we take e.r. ~ $0.7\times10^4$ W/m$^3$ (7kW/m$^3$), and I(0) ≈ $15\times10^5$ W/m$^3$ (1.5 MW/m$^3$), the result is l ≅ 7.36 m.

Thus $T_1$ can be stabilized (see Appendix at the end) because $T_2$ remains constant during the boiling of the water (of course the steam so produced is the one which will make the turbine to operate in order for electricity to be finally produced).

**Appendix**

In order to examine the stability of the temperature $T_1 \equiv T$ inside the torus, we find from eqn. (12.7) (setting $T_2 \equiv T'$)

$$\frac{T-T'}{I(0)+e.r.} = \frac{r_0^2}{2\kappa}\ln\left(1+\frac{l}{r_0}\right), \qquad \text{(A1)}$$

or

$$e.r. = \frac{2\kappa}{r_0^2}\ln^{-1}\left(1+\frac{l}{r_0}\right)(T-T') - I(0), \qquad \text{(A2)}$$

so that, for small T′, we will have

$$(e.r.)_0 \cong \frac{2\kappa}{r_0^2}\ln^{-1}\left(1+\frac{l}{r_0}\right)T_0 - I(0) \qquad \text{(A3)}$$

at equilibrium, and a small change δ(e.r.) of e.r. will cause a corresponding change δT in T (T′ will of course remain constant at 100° K) given by

$$\frac{\delta\ (e.r.)}{\delta\ T} = \frac{2\kappa}{r_0^2} \ln^{-1}\left(1+\frac{l}{r_0}\right). \tag{A4}$$

In our case we take $r_0$ = 100 cm, l = 736 cm, κ = 0.0004 cal $sec^{-1}$ $cm^{-1}$ $grad^{-1}$, and $T_0 = 10^7$ °K. We then find from (A3) that $(e.r.)_0 \cong 7$ kW/$m^3$, as we have already seen, and from (A4) δ(e.r.)/δT = 0.15 (W/$m^3$) $deg^{-1}$.

But we can find the derivative d(e.r.)/dT from the function e.r. = f(T), since in reality e.r. depends on the temperature (see Fig.5). In fact, from eqns. (8.4) and (8.2′) we have

$$\frac{e.r.}{1.44MeV} = \frac{1}{2} N^2 \sigma_{pp} \sqrt{\frac{3kT}{m_+}}, \tag{A5}$$

or

$$e.r. = \frac{1.44MeV}{2} N^2 \sigma_{pp} \sqrt{\frac{3k}{m_+}} \sqrt{T}, \tag{A6}$$

so that, at equilibrium,

$$(e.r.)_0 = \frac{1.44MeV}{2} N^2 \sigma_{pp} \sqrt{\frac{3k}{m_+}} \sqrt{T_0}. \tag{A7}$$

Thus, differentiating (A6) we find

$$\frac{d(e.r.)}{dT} = \frac{1.44MeV}{4} N^2 \sigma_{pp} \sqrt{\frac{3k}{m_+}} T^{-1/2}. \tag{A8}$$

In our case we find from (A7), and for $\sigma_{pp} \sim \sigma_{pe} \sim 10^{-25}$ $cm^2$, that N ≅ $10^{14}$ $cm^{-3}$. But the value $\sigma_{pp} \sim \sigma_{pe}$ is unrealistic. In fact [9] the cross-section is given by

$$\sigma\ (E) = \frac{S(E)}{E} e^{-b/\sqrt{E}}, \tag{A9}$$

where E = kT. In our case we have E ≅ 1 keV, so that S(E) ~ 1 keV barn. Also b is given by [9]

where $Z_1$ and $Z_2$ are the atomic numbers (here $Z_1 = Z_2 = 1$) and A the reduced atomic weight, given by

where E = kT. In our case we have E ≈ 1 keV, so that S(E) ~ 1 keV barn. Also b is given by [9]

$$b = 31.28 \cdot Z_1 Z_2 \sqrt{A}\ \sqrt{keV}, \qquad \text{(A10)}$$

where Z1 and Z2 are the atomic numbers (here Z1 = Z2 = 1) and A the reduced atomic weight, given by

$$A = \frac{A_1 A_2}{A_1 + A_2} \qquad \text{(A11)}$$

(here A = ½). We find in our case, from (A10), that b = 22.2 √(keV). We thus finally find from (A9) that $\sigma_{pp}(10^7\ °K) \cong 2\times10^{-10}$ barn $= 2\times10^{-34}$ cm$^2$. Now, from (A7) we see that $N^2\sigma_{pp}$ is constant. And, since for $\sigma_{pp} \sim 10^{-25}$ cm$^2$ we have found above that $N \cong 10^{14}$ cm$^{-3}$, we see that for $\sigma_{pp} \cong 2\times10^{-34}$ cm$^2$ we must take $N \cong 2.2\times10^{18}$ cm$^{-3}$. With these values of $\sigma_{pp}$ and N, along with $T = T_0 = 10^7$ °K, we find thus finally from (A8) that d(e.r.)/dT $\cong 3.5\times10^{-4}$ (W/m$^3$) deg$^{-1}$.

Now it is obvious that if

$$\frac{\delta\ (e.r.)}{\delta\ T} > \frac{d(e.r.)}{dT} = f_0' \qquad \text{(A12)}$$

at $T = T_0$ for a small increment of e.r., then the equilibrium at $T = T_0$ is stable (see Fig.5). Otherwise it is unstable. Hopefully, we can always secure inequality (A12) by appropriately varying the parameters of our problem. Here inequality (A12) largely holds, since δ(e.r.)/δT = 0.15 (W/m$^3$) deg$^{-1}$ and d(e.r.)/dT = 0.00034 (W/m$^3$) deg$^{-1}$.

Note here that δ(e.r.)/δT depends only on the reactor characteristics, while d (e.r.)/dT depends on the reactions involved (cf. A4 and A8 respectively).

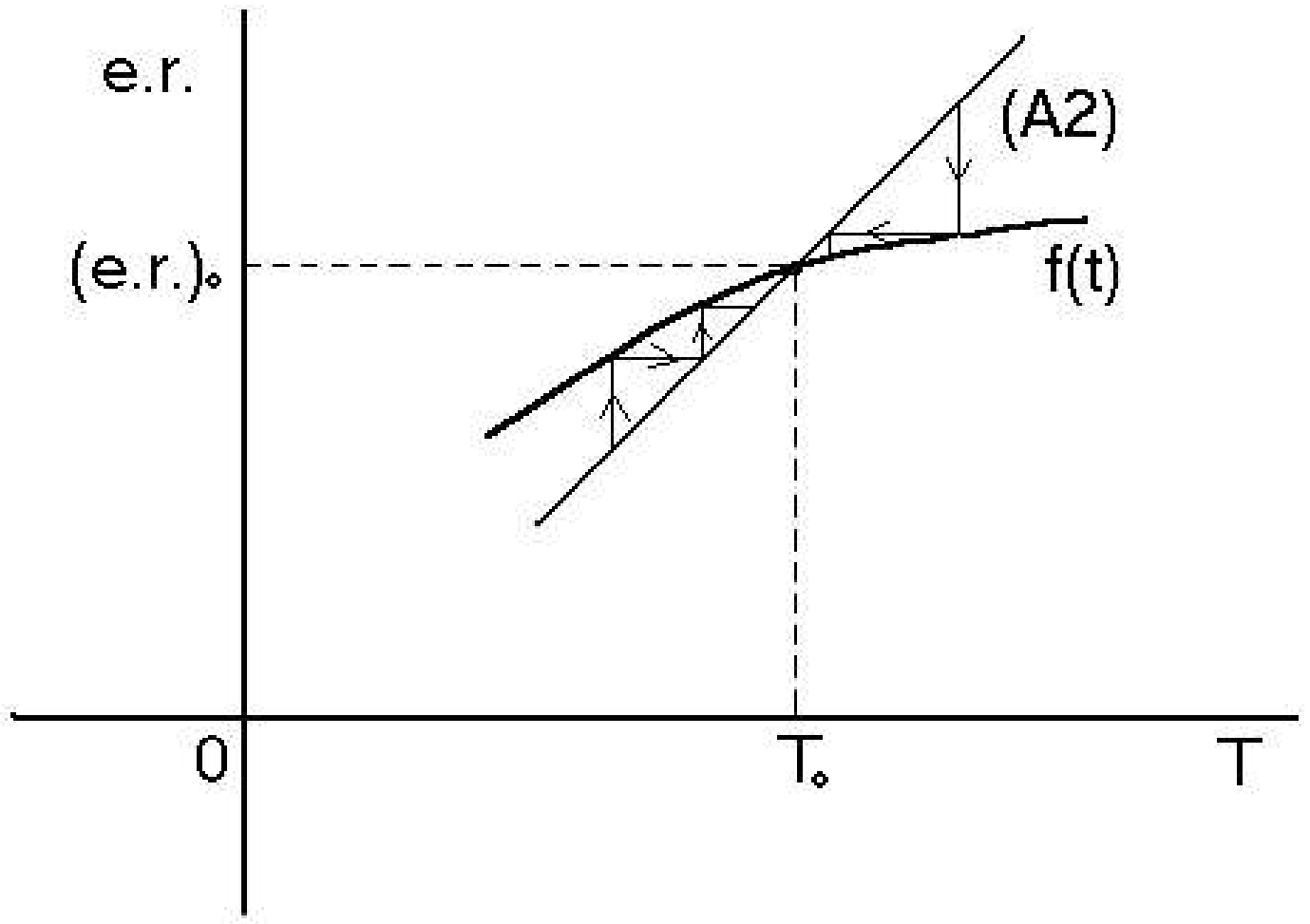

Fig.5

**Comment**

In §12 and in the Appendix we used a temperature $T_1$ as indicating the temperature of the plasma. But our (almost) collisionless plasma is characterized by two temperatures: $T \equiv T_+$ & $T' \equiv T_- = 1836\ T_+$ . Nevertheless, collisions certainly occur between protons and electrons at a density of ~ $10^{18}$ $cm^{-3}$ , with the result for the two temperatures to converge *(slowly)* to a unique tempetarure between them. Quoting [14]: "The large difference between the electron mass m and the ion mass M impedes the transfer of energy between electrons and ions: when a heavy and a light particle collide, the energy of each is almost unchanged. The establishment of equilibrium among the electrons alone an the ions alone therefore takes place considerably more quickly than that between electrons and ions. This may easily give rise to a situation where the electron and ion components of the plasma each have a Maxwelllian distribution but with different temperatures $T_e$ and $T_i$ , of which $T_e$ is usually the greater.”

The equations governing the change of T and T′ are [14]

$$\frac{dT}{dt} = \frac{T' - T}{\tau} \tag{C1}$$

and

$$\frac{dT'}{dt} = -\frac{T' - T}{\tau}, \tag{C2}$$

where τ is a characteristic (relaxation) time for the establishment of proton-electron thermal equilibrium.

We see at once from these two equations that their right-hand sides are opposite. (This is simply implied by the law of conservation of energy.) This results in the equation

$$\frac{dT}{dt} + \frac{dT'}{dt} = 0. \tag{C3}$$

Thus we obtain

$$T + T' = \text{ constant } \equiv C(= 1837 \cdot T_+), \tag{C4}$$

so that it is evident that

$$T_{\lim} = C/2. \tag{C5}$$

If we want to find explicitly the dependence of T (& T′) on the time t, we have then to solve the equation

$$\tau \ \dot{T} + 2T = C. \tag{C6}$$

This equation results from the foregoing, and it is a linear ordinary differential equation of the first order (with constant coefficients). Its solution is

$$T = Ge^{-\frac{2}{\tau}t} + \frac{C}{2}, \tag{C7}$$

with G another constant given by

$$G = -\frac{T_- - T_+}{2}. \tag{C8}$$

**Remark**

If $S_0$ is the outer surface of the plasma and $T_0$ its temperature, and $S_1$ is the inner surface of the covering concave torus with $T_1$ the temperature at the location of $S_1$, it is meant that $S_0 = S_1$ and $T_0 = T_1$ in the main text. But in reality this cannot be the case. In fact, no material can exist at the temperature of the plasma without melting. Thus we have to consider the transfer of heat power Q/t from $S_0$ to $S_1$ by radiation. We have then from the Stefan-Boltzmann law [3] that

$$Q/t = \sigma\left(S_0 T_0^4 - S_1 T_1^4\right). \qquad \text{(R1)}$$

Letting

$$Q/t = [I(0) + (e.r.)]\,V_0 \qquad \text{(R2)}$$

we can find the actual $T_1$, from the combination of (R1) with (R2), which of course must be much less than $T_0 \sim 10^7$ °K. It can be so possible for the material torus to consist of a ceramic material, for example.

We remind here that, if we consider a torus of small radius r and big radius R, its surface S and its volume V are given by

$$S = 4\pi^2 Rr \qquad \text{(R3)}$$

and

$$V = 2\pi^2 Rr^2. \qquad \text{(R4)}$$

Thus (R1) and (R2) can be written

$$Q/t = 4\pi^2 R\sigma\left(r_0 T_0^4 - r_1 T_1^4\right) \qquad \text{(R5)}$$

and

$$Q/t = 2\pi^2 R r_0^2 [I(0) + (e.r.)], \qquad \text{(R6)}$$

so that , equating the right-hand sides,

$$2\sigma\left(r_0 T_0^4 - r_1 T_1^4\right) = r_0^2 [I(0) + (e.r.)]. \qquad \text{(R7)}$$

$$r_1 T_1^4 = r_0 T_0^4 - \frac{r_0^2}{2\sigma}\left[I(0) + (e.r.)\right]. \qquad \text{(R8)}$$

We set $r_0$ = 1 m, $T_0$ = $10^7$ °K, σ = 5.7×$10^{-8}$ W/($m^2$grad$^4$). Then, even if we take I (0) = 3.3×$10^{10}$ W/$m^3$ and (e.r.) = 3.4×$10^{12}$ W/$m^3$, we find $r_1T_1^4$ = $10^{28}$ m grad$^4$, so that for $T_1$ = 1500 °K we obtain $r_1$ = 2×$10^{15}$ m ! Thus, we have to (at last analysis) increase N much more (than N = 2.2×$10^{18}$ cm$^{-3)}$ in order to increase I(0) (∝ N) and, in particular, (e.r) (∝ $N^2$).

In practice it must be

$$\frac{r_0^2}{2\sigma}\left[I(0) + (e.r.)\right] \approx r_0 T_0^4 \qquad \text{(R9)}$$

(of the same order). Then we find that, since I(0) + (e.r.) is of the same order to (e.r.) (I(0) ~ 0), we must take (e.r.) ~ 1.14×$10^{21}$ W/$m^3$ (*). If we increase N by a factor of $10^4$ (corresponding to N = 2.2×$10^{22}$ cm$^{-3}$), we have to multiply I(0) by $10^4$, and (e.r.) by $10^8$. We take thus I(0) = 3.3×$10^{14}$ W/$m^3$ and (e.r.) = 3.4×$10^{20}$ W/$m^3$. (§)

We know that [11] the most attractive reactions for use in a thermonuclear fusion reactor appear to be the D-D reaction

$$H^2(d,n)He^3, \qquad Q = 3.27 MeV \qquad \text{(R10)}$$

$$H^2(d,p)H^3, \qquad Q = 4.03 MeV \qquad \text{(R11)}$$

and the D-T reaction

$$H^3(d,n)He^4, \qquad Q = 17.59 MeV \qquad \text{(R12)}$$

The total heat produced by these reactions is 24.89 MeV, that is 17.28 times greater than 1.44 MeV used up to now. Thus, if we multiply (§) by the same factor (17.28), we obtain (e.r.) = 5.9×$10^{21}$ W/$m^3$ ($), exceeding the value (*) required. In other words we can increase (e.r.) to the necessary value without increasing N more, by just taking account of the actual reactions taking place in the reactor.